\documentclass{article}
\usepackage{spconf, amsmath,graphicx}
\usepackage{amssymb}
\usepackage{amsmath}
\usepackage[fleqn]{nccmath} 
\usepackage{graphicx} 
\usepackage{IEEEtrantools}
\usepackage{caption}
\usepackage{multirow,multicol}
\usepackage{subcaption}
\usepackage{comment}
\usepackage{bbm}

\makeatletter
\newcommand{\thickhline}{%
    \noalign {\ifnum 0=`}\fi \hrule height 1pt
    \futurelet \reserved@a \@xhline
}
\title{Learning domain invariant representations for child-adult classification from speech}
\name{Rimita Lahiri$^1$, Manoj Kumar$^1$, Somer Bishop$^2$, Shrikanth Narayanan$^1$}
\address{
  $^1$Signal Analysis and Interpretation Laboratory, University of Southern California\\
  $^2$Department of Psychiatry, University of California, San Francisco}

\begin{document}
%
\maketitle
\begin{abstract}
Diagnostic procedures for ASD \textit{(autism spectrum disorder)} involve semi-naturalistic interactions between the child and a clinician. Computational methods to analyze these sessions require an end-to-end speech and language processing pipeline that go from raw audio to clinically-meaningful behavioral features. An important component of this pipeline is the ability to automatically detect who is speaking when i.e., perform child-adult speaker classification. This binary classification task is often confounded due to variability associated with the participants' speech and background conditions.
Further, scarcity of training data often restricts direct application of conventional deep learning methods. In this work, we address two major sources of variability--age of the child and data source collection location--using domain adversarial learning which does not require labeled target domain data.
We use two methods, generative adversarial training with inverted label loss and gradient reversal layer to learn speaker embeddings invariant to the above sources of variability, and analyze different conditions under which the proposed techniques improve over conventional learning methods. 
Using a large corpus of \textit{ADOS-2 (autism diagnostic observation schedule, 2nd edition)} sessions, we demonstrate upto 13.45\% and 6.44\% relative improvements over conventional learning methods.

\end{abstract}
\section{Introduction}
\label{sec:intro}

Autism spectrum disorder (ASD) refers to a group of neuro- developmental disorders characterized by abnormalities in speech and language \cite{volden1991neo,languageAsd2014Kim,huemer2010comprehensive} and often diagnosed in children using semi-structured dyadic interactions with a trained clinician. The reported ASD prevalence has been steadily increasing among children in the US: from 1 in 150 \cite{centers2006mental} to 1 in 59 \cite{baio2018prevalence}, 
Computational processing of the participants' speech and language during such child-adult interactions has shown potential in recent years in supporting and augmenting human perceptual and decision making capabilities. \cite{bone2016use,bone2016acoustic,kumar2016objective}.

\begin{figure}
    \centering
    \includegraphics[width=0.5\textwidth]{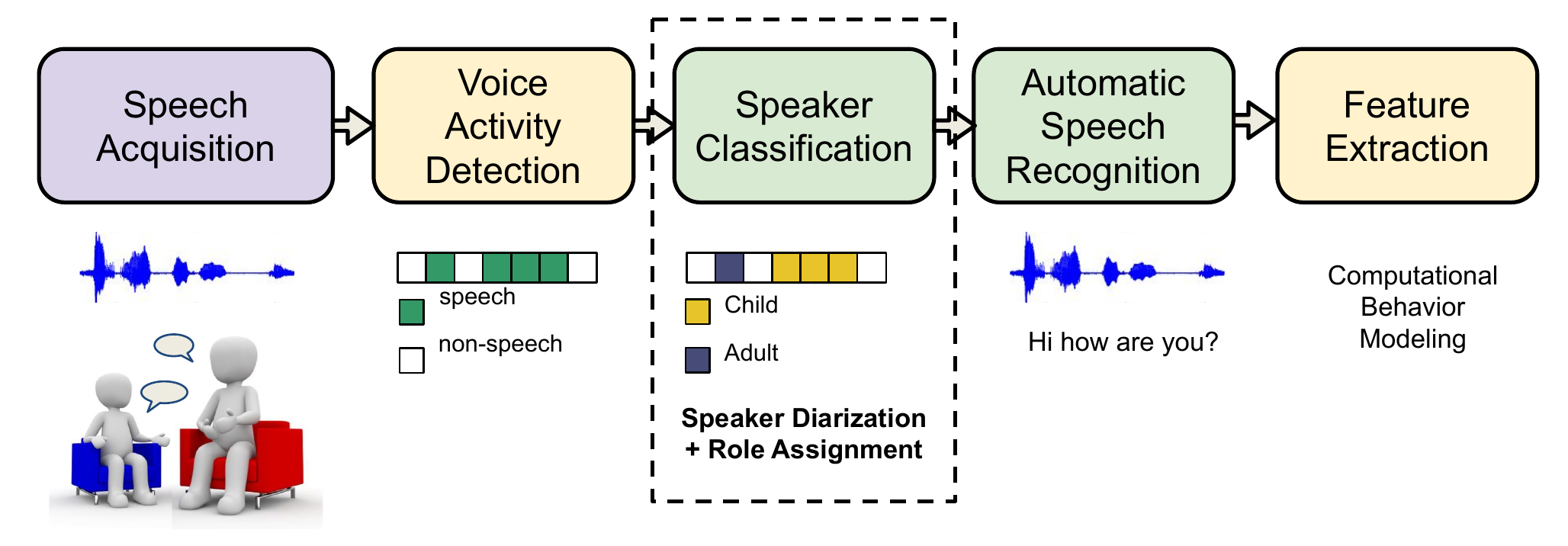}
    \vspace{-3.5ex}
    \caption{Speech processing pipeline for feature extraction}
    \label{fig:care_pipeline}
\vspace{-3.5ex}
\end{figure}

However, previous works utilized manual speaker labels and transcripts for behavioral feature computation, which can be expensive and time-consuming to create. Hence, feature extraction at-scale is dependent on a robust speech and language pipeline (Figure \ref{fig:care_pipeline}). An important component of the pipeline is speaker diarization, which answers the question ``\textit{who spoke when?}". In the context of ASD diagnostic assessment sessions, diarization can be approached as (supervised) child-adult speaker classification. Training a child-adult classification system is often not straightforward due to multiple sources of variability in the data. Among others, two primary sources of variability arise from developmental aspects of child speech \cite{lee1999acoustics} and from varying background conditions, often influenced by where and how the data are collected. In this work, we train a child-adult classification system using domain adversarial training \cite{wilson2019survey,ganin2016domain} to address these sources of variability.

\begin{figure}
\centering
\begin{subfigure}[b]{0.9\textwidth}
   \includegraphics[scale=0.2]{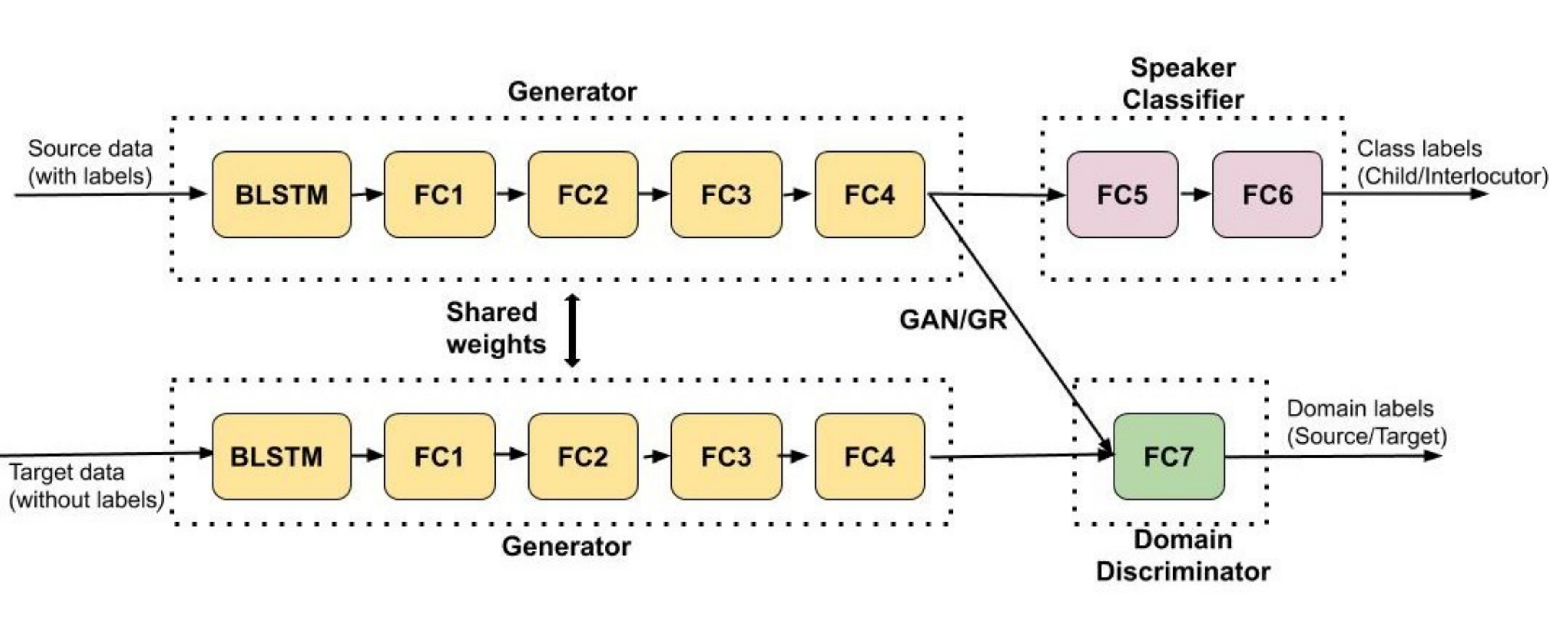}
\end{subfigure}
\begin{subfigure}[b]{0.9\textwidth}
   \includegraphics[scale=0.2]{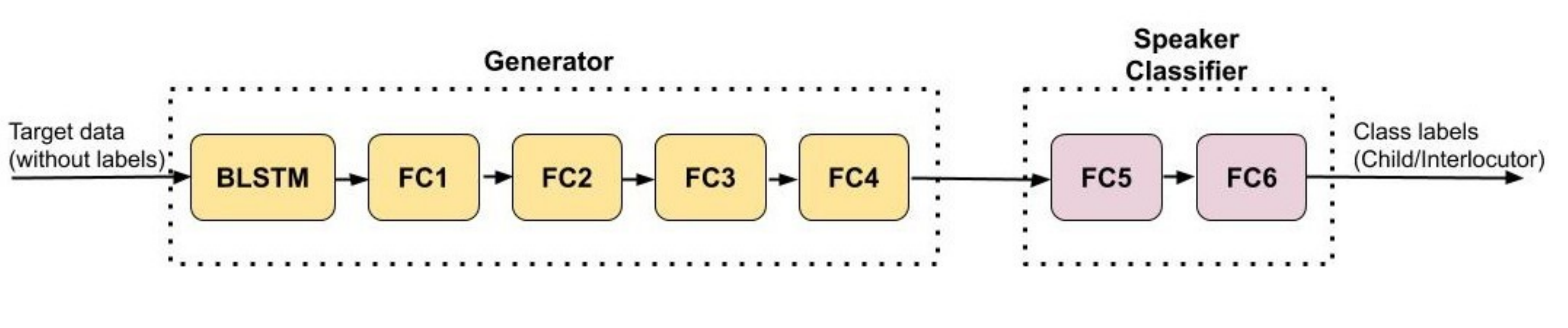}
\end{subfigure}
\vspace{-4.0ex}
\caption{Training and Testing Network Architecture}
\label{fig:main_network}
\vspace{-3.0ex}
\end{figure}

A generative adversarial network (GAN) is composed of two mutually pitting neural networks, termed as the generator and the discriminator. These networks play a minimax game, where the generator aims to create fake samples from a noise vector of some arbitrary distribution in order to confuse the discriminator. On the other hand, the discriminator tries to distinguish between the real and fake samples. 
Domain adversarial learning can be formulated as a variant of GANs, where the noise vectors are replaced with target data, and the (domain) discriminator network tries to discriminate whether a sample belongs to source or target domain. Hence, the generator network learns to extract domain-invariant representations. The speaker classifier is trained on the generator outputs in a multi-task manner.
In this work, we have used two different methods of domain adversarial training namely \textit{Gradient Reversal (GR)} \cite{ganin2016domain} and \textit{Generative Adversarial Networks (GAN)} \cite{bhattacharya2019adapting}. GR tries to learn the domain-invariant feature by reversing the gradients coming from domain discriminator while GAN aims to achieve the same by training with inverted domain labels. The full network configuration comprising of generator (feature extractor), discriminator and speaker classifier is shown in Figure \ref{fig:main_network}.


The rest of the paper is organized as follows: Section \ref{sec:background} provides a brief overview of the background works. Section \ref{sec:methods} describes the domain adversarial methods used in this work. Section \ref{sec:expt} provides experimental details and details of the dataset used. Key outcomes of the experiments are tabulated and interpreted in section \ref{sec:results}. Finally, section 6 provides conclusions and highlights possible future extensions.

\section{Background}
\label{sec:background}
\subsection{Speaker Diarization in Autism Diagnosis Sessions}

Although there exists a significant amount of work in speaker diarization of broadcast news and meetings, interest in spontaneous and real-life conversations has emerged only recently. Diarization solutions for child speech (both child-directed and adult-directed) initially looked at traditional feature representations (MFCCs, PLPs) \cite{najafian2016speaker} and speaker segmentation/clustering methods (generalized likelihood ratio, Bayesian information criterion) \cite{zhou2016speaker,sun2018}. In \cite{zhou2016speaker}, the authors introduced several methods for working with audio collected from children with autism using a wearable device. More recently, approaches based on fixed-dimensional embeddings such as ivectors \cite{cristia2018talker} and DNN speaker embeddings such as x-vectors \cite{xiel2019} were explored. 
While some of the above approaches have adapted clustering methods to child speech \cite{xiel2019}, to the best of our knowledge none of them have taken into account shifts in domain distribution that is likely to adversely impact diarization performance.

\subsection{Domain Adversarial Learning}
Domain adaptation within adversarial learning was first introduced by \cite{ganin2016domain} for computer vision related applications. Since then, there has been an emerging trend to use domain adversarial learning to alleviate the mismatch between the training and testing data in various speech applications including ASR and acoustic emotion recognition \cite{abdelwahab2018domain}. In \cite{tripathi2018adversarial,sun2018domain} the authors have employed domain adversarial training to improve the robustness of the speech recognition system to handle different noise types and levels. In \cite{denisov2018unsupervised}, the authors applied domain adversarial training to address mismatch between close-talk and single-channel far-field recordings.
Our motivation for applying domain adversarial learning is inspired from recent applications (\cite{bhattacharya2019adapting,bhattacharya2019generative}) in speaker verification across multiple languages. It was shown that adversarial training can be used to learn robust speaker embeddings across different conditions. We extend this concept to the task of child-adult classification from speech, where 
variabilities in children's linguistic capabilities and recording locations can be viewed as domain shift that can be modeled using adversarial learning.

\section{Domain Adversarial Learning for Speaker classification}
\label{sec:methods}

The main aim of the work is to efficiently distinguish between the speakers (namely, child and an adult interlocutor) from audio recordings of diagnostic sessions from different clinical locations. Besides learning domain invariant features by confusing the discriminator, the network must be able to efficiently distinguish between the speakers as well. In this work, we have shown that the proposed objective can be accomplished using a GAN based method, or a GR based method. 

Consider samples from the source domain $(X_s,Y_s) \in \Omega_s$ and target domain $(X_t, Y_t) \in \Omega_t$ with a common label space $Y$. During training, labels from the target domain are assumed unavailable, and data distributions of $X_s$ and $X_t$ might differ. The goal of domain adversarial learning is to maximize the target accuracy by jointly learning to maximize task performance and reducing domain shift between the source and target domains in generator output embedding space.

In our work, we begin by training the network with source data and corresponding speaker labels to minimize task loss.
We refer to this as \textit{pre-training}. Following, the adversarial game continues where the discriminator is trained with true domain labels and the generator is trained either with inverted domain labels \textit{(GAN)} or reverse gradients \textit{(GR)} alternatively until convergence is reached.

In both methods, for every batch of data, the training is carried out in three distinct steps. In the first step, the generator and speaker classification models are trained with true speaker labels from the source data using the following objective:

\begin{fleqn}
\begin{equation}
\begin{aligned}
 &\hspace{6.0ex} \min\limits_{G,C}Loss_{Spk}(X_s,Y_s)= \\
 &\hspace{6.0ex} \mathop{\mathbb{E}_{x_s,y_s \sim (X_s,Y_s)}\sum_{k=1}^{2}\mathbbm{1}_{k=y_s}\log(C(G^{s}(x_{s})))}
\end{aligned}  
\end{equation}
\end{fleqn}

\noindent where $G(.)$ and $C(.)$ are the generator and classifier functions, respectively.
In the second step, the embeddings are extracted from the output layer of the generator for both source and target data using the model trained in the previous step. The domain discriminator is now trained with the true domain labels. This step ensures that the discriminator is well trained to distinguish between source and target domain.
\begin{fleqn}
\begin{equation}
\begin{aligned}
    \min\limits_{D}Loss_{Dom}(X_{s},X_{t},G)=\mathop{\mathbb{E}}_{x_{s} \sim X_s}\log(D(G(x_s)))+\\ \mathop{\mathbb{E}}_{x_{t}\sim X_{t}}\log(1-(D(G(x_{t}))))
\end{aligned}  
\end{equation}
\end{fleqn}
The first and second steps are the same for both GAN and GR: they differ in the third step. For GAN, the generator is trained with source and target data but with inverted domain labels: 
\begin{fleqn}
\begin{equation}
\begin{aligned}
    \min\limits_{G}Loss_{Adv}(X_{s},X_{t},G)=\mathop{\mathbb{E}}_{x_{s} \sim X_s}log(D(G(x_t)))+\\\mathop{\mathbb{E}}_{x_{t} \sim X_{t}}log(1-(D(G(x_{s}))))
\end{aligned}  
\end{equation}
\end{fleqn}
In case of GR, the gradients from the domain discriminator are reversed for training. In both the cases, the final step ensures the generator is trained well to generative domain-invariant representations. It is important to note that the generator network weights are updated twice during the adversarial training in first and last step respectively. 

\begin{figure*}[!t]
\captionsetup[subfigure]{justification=centering}

    \begin{subfigure}[b]{0.24\textwidth}
        \includegraphics[width=\textwidth]{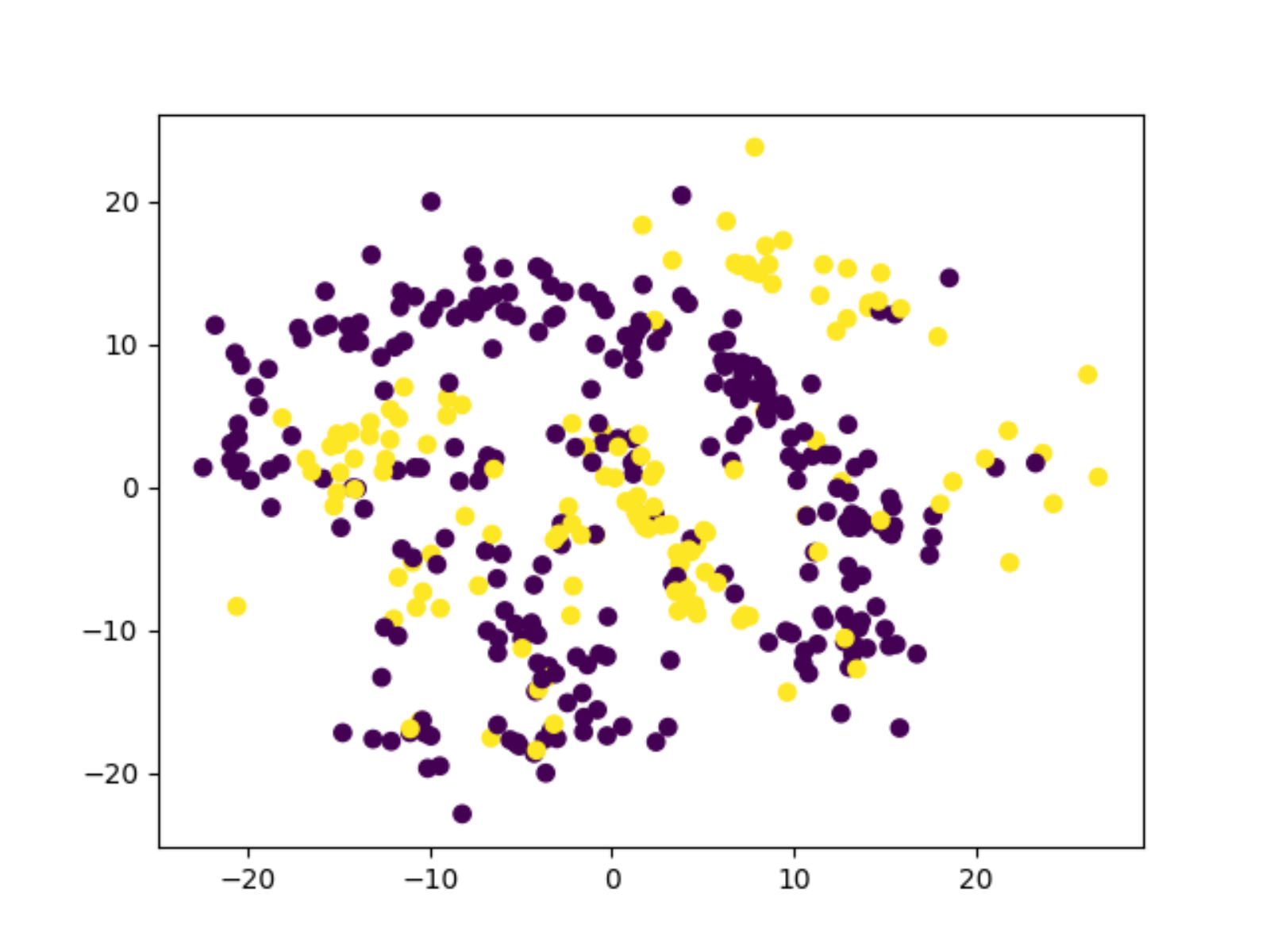}
        \caption{Session A:\\ Before adversarial training}\label{fig1.a}
    \end{subfigure}
    \begin{subfigure}[b]{0.24\textwidth}
        \includegraphics[width=\textwidth]{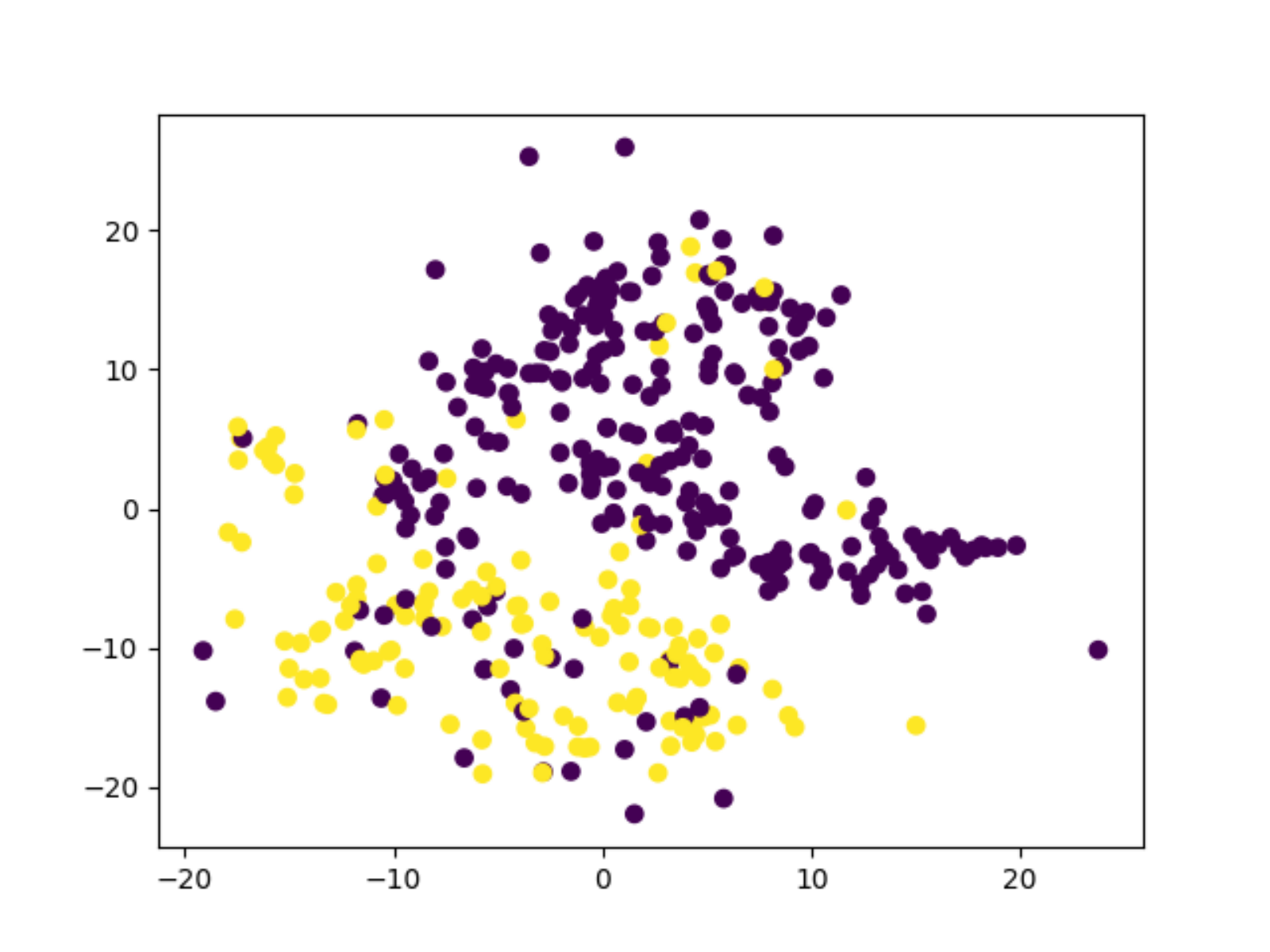}
        \caption{Session A:\\ After adversarial training}\label{fig1.b}
    \end{subfigure}
    \begin{subfigure}[b]{0.24\textwidth}
        \includegraphics[width=\textwidth]{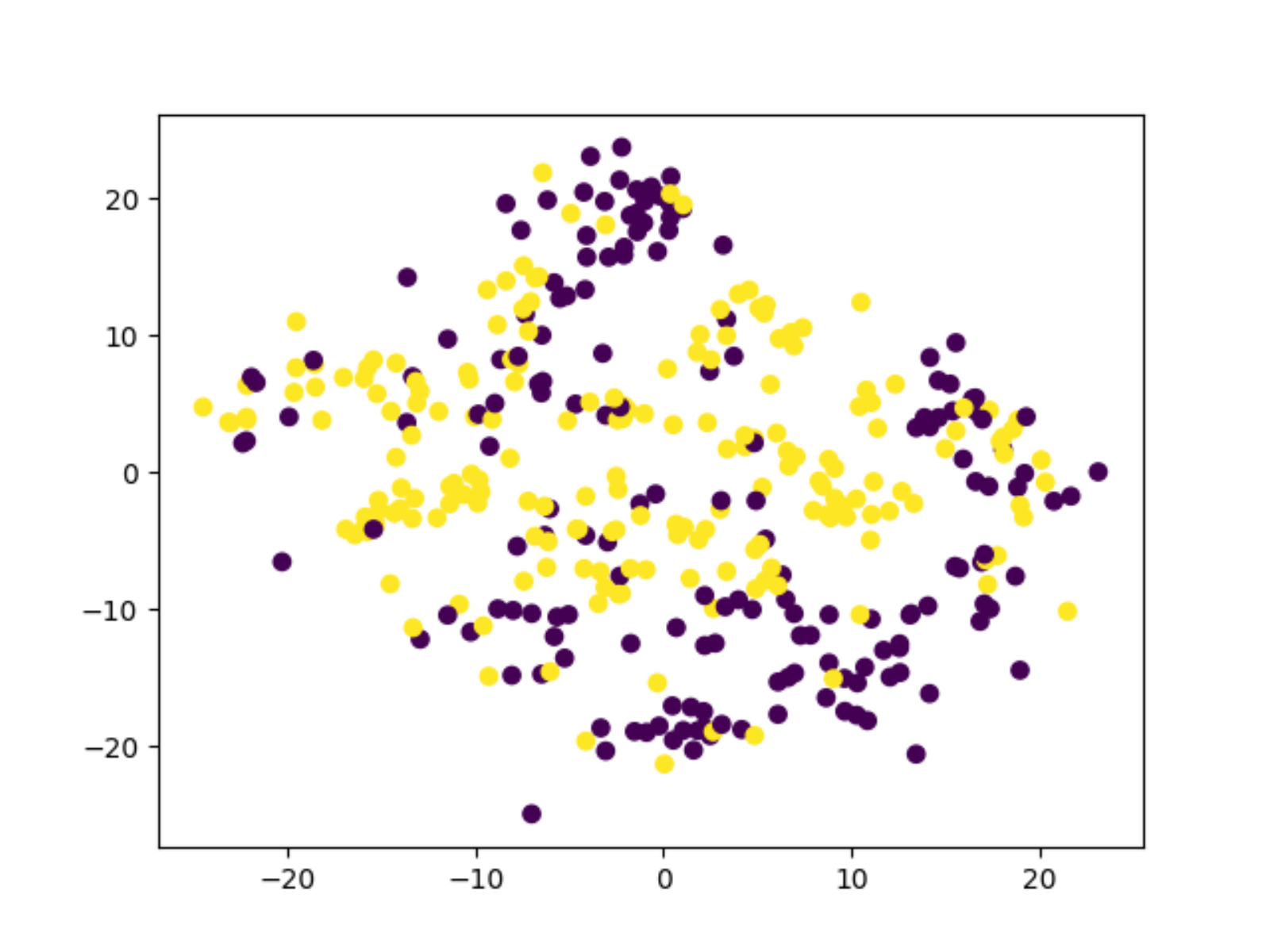}
        \caption{Session B:\\ Before adversarial training}\label{fig1.c}
    \end{subfigure}
    \begin{subfigure}[b]{0.24\textwidth}
        \includegraphics[width=\textwidth]{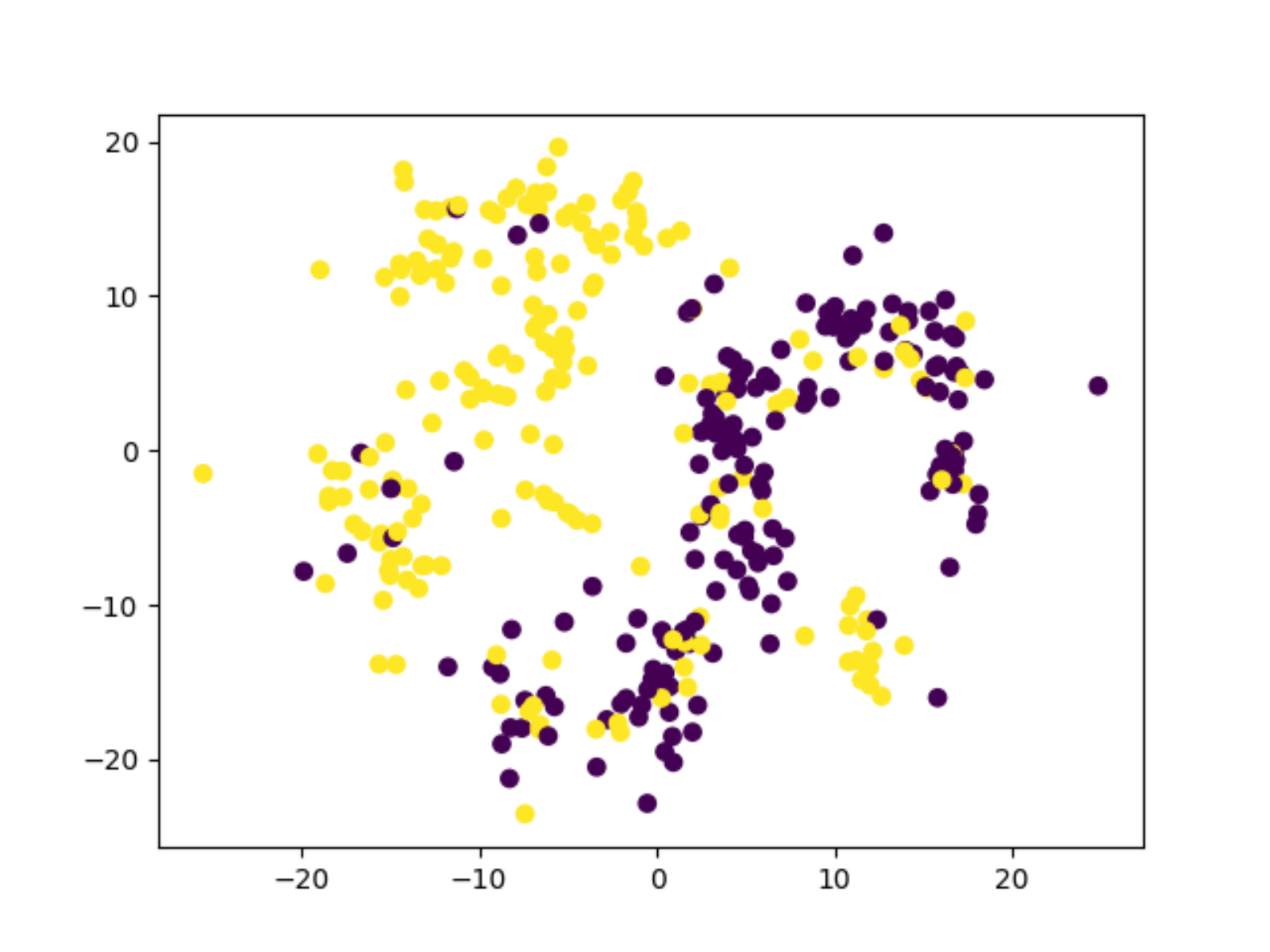}
        \caption{Session B:\\ After adversarial training}\label{fig1.d}
    \end{subfigure}
    \vspace{-1.5ex}
    \caption{TSNE plots of the most discriminative 2 components of the generator output corresponding to the classes}
    \label{fig:tsne}
\vspace{-2.5ex}
\end{figure*}

\section{Experimental setup} 
\label{sec:expt}

\subsection{Dataset} 

The ADOS-2 dataset is composed of semi-structured activities involving a child and an interlocutor, who is trained to examine behaviours related to ASD.
A typical ADOS-2 session lasts between 40-60 minutes and consists of varying subtasks designed to elicit responses from the child under different social and interactive circumstances. 
In this work, we look at administrations of Module-3 which are intended for verbally-fluent children. Further, we restrict to the \textit{Emotions} and \textit{Social Difficulties \& Annoyance} subtasks since they elicit spontaneous speech from the child under significant cognitive load.
In the \textit{Emotions} subtask the child is asked to recognize different objects that trigger various emotions within them and share their perceptions on the same. 
The \textit{Social Difficulties \& Annoyance} subtask explores the child's thoughts regarding various social problems faced at home or school. 
The dataset consists of recordings from 165 children (86 ASD, 79 Non-ASD) collected from two different clinical centers: University of Michigan Autism and Communciation Disorders Center (UMACC) and Cincinnati Children's Medical Center (CCHMC). Further details are presented in Table \ref{tab:dataset}.

\begin{table}[t!]
\begin{center}
\caption{Demographic details of ADOS dataset}
\label{tab:dataset}
\resizebox{0.48\textwidth}{!}{%
 \begin{tabular}{c c} \thickhline
 \textbf{Category} & \textbf{Statistics}  \\  \thickhline
 Age(years) & Range: 3.58-13.17 (mean,std):(8.61,2.49)  \\  \hline
 Gender & 123 male, 42 female \\ \hline
 Non-verbal IQ & Range: 47-141 (mean,std):(96.01,18.79) \\  \hline
  \multirow{4}{4em}{\centering Clinical Diagnosis} & 86 ASD,42 ADHD \\  
 & 14 mood/anxiety disorder\\
 & 12 language disorder\\
 & 10 intellectual disability, 1 no diagnosis\\  \hline
 \multirow{2}{4em}{\centering Age distribution} & Cincinnati: $\leq$5yrs 7, 5-10 yrs 52, $\geq$10yrs 25 \\
 & Michigan: $\leq$5yrs 11, 5-10 yrs 42, $\geq$10yrs 28\\ \thickhline
\end{tabular}}
\end{center}
\vspace{-4.5ex}
\end{table}

\subsection{Features and neural network architecture}

In all experiments we used 23-dimensional MFCC features with mean and variance normalized at session level. The features were extracted using the Kaldi\footnote{https://github.com/kaldi-asr/kaldi} toolkit with a frame-length of 40ms and frame-shift of 20ms. Features were spliced with a context of 15 frames yielding a sample of dimension 31$\times$23. Consecutive samples were chosen with an interval of 15 frames in order to minimize overlap during DNN training. \par
The generator $G(.)$ consists of a bidirectional long short term memory (BLSTM) layer followed by four dense layers consisting of 128, 64, 16 and 16 neurons respectively.
Certain settings have smaller training data compared to others, hence the number of parameters were reduced to prevent over-fitting.
The \textit{speaker classifier} $C(.)$ consists of two dense layers with 16 neurons each, while the \textit{domain discriminator} $D(.)$ consists of one dense layer of 16 neurons. Rectified linear units (ReLU) layers were used as activation functions for all the layers, and both dropout ($p$ = 0.2) and batch normalization were applied to every hidden layer for regularization.

\subsection{Baselines}

We have compared the performance of our systems with two systems. The first system (\textit{Pre-Train}) is composed of only the feature generator and the speaker classifier blocks. This system is trained with source data and directly tested on target data, the goal being to check whether domain adversarial training provides any improvement over pre-training.
The second model uses the same architecture, except the training data is augmented with target domain data. Since target labels are not available during domain adversarial training, this system (\textit{Upper-Bound})serves as an upper bound for the performance.

\subsection{Cross-Domain Design}
To address the variability resulting from child age and location differences, we designed two sets of experiments: 
First, we partitioned the data according to age groups and chose the two farthest groups from both locations as the source and the target data. In (\textit{Exp 1}), we selected sessions of kids (\textit{$\geq$10yrs}) as the source and sessions of kids (\textit{$\leq$5yrs}) as target data. Later in (\textit{Exp 2}), we reversed the source and target data and repeated the same experiment to address domain shift in the other direction.\par
Second, we divided the sessions based on their locations. To control for variability sources, we further divided the sessions from each location into 3 age groups of (\textit{$\leq$5yrs}, \textit{5-10yrs}, \textit{$\geq$10yrs}) and conducted separate experiments within each group. In (\textit{Exp 3}),  for each age group we considered recordings from Cincinnati as source data and recordings from Michigan as target data. Later, in \textit{Exp 4} we reversed the source and target data and conducted the same experiment).\par
We check for complementary information in embeddings extracted from GAN and GR using score fusion and embedding fusion. 
For the score fusion system, we estimate class distribution for a test sample by computing posterior means from GAN and GR models.  
For the embedding fusion system, we extract embeddings from the output of the generator block for both source and target data for GAN and GR. We then concatenate GAN and GR embeddings and train a separate neural network model with similar architecture to the GAN and GR models, using the source data. Finally, the fused embeddings of the target data are fed to the trained network to check classification performance.\par
For all experiments, we update model weights using Adam optimizer (lr = 0.001, $\beta_1$ = 0.9, $\beta_2$ = 0.999, $\epsilon$ = $1\mathrm{e}{-8}$) to minimize categorical cross-entropy loss. Accuracy on a set of held-out sessions from the source corpora is used for early stopping during both pre-training and domain adversarial training. 
During evaluation, we discard the domain discriminator part. The 23-dimensional features from the audio session are fed to the network consisting of the \textit{generator} $G(.)$ and the \textit{speaker classifier} $C(.)$ to estimate speaker labels at sample-level. 
Since many sessions in our corpus contain imbalanced class distributions (more samples from adult than child), we estimate classification performance using the mean (unweighted) F1-score.


\begin{table}[t!]
\centering
\caption{Mean F1-score (\%) treating child age as domain shift}
\label{tab:age_expt}
\begin{tabular}{ccc} \thickhline
\textbf{Systems} & \textbf{Exp 1(\%)} & \textbf{Exp 2(\%)}\\ \thickhline
Pre-Train & 73.40  & 63.69\\
GAN &  78.27 & 71.21\\
GR & 78.53 & \textbf{72.26} \\
Score Fusion & \textbf{78.86} & 71.61   \\
Embed. Fusion & 78.38 & 71.95   \\ \hline
Upperbound & 85.65 & 86.29   \\\thickhline
\end{tabular}
\vspace{-1.5ex}
\end{table}

\section{Results and Analysis} 
\label{sec:results}

From Tables \ref{tab:age_expt} and \ref{tab:location_expt}, we observe that both GAN and GR outperform the baselines in age-based and location-based experiments. In general, GR performs slightly better than GAN in a majority of settings.
Among the age-based experiments, we observe that Exp 2 which consists of kids aged $\geq$10yrs as target data, degrades in accuracy for all models. A possible reason is that older kids with well-developed vocal tract and speaking skills are harder (i.e., more confusable) for the model to discriminate from adult speakers. Interestingly, domain adaptation returns a greater relative improvement over pre-training in Exp 2 (13.45\%) than Exp 1 (7.43\%). 

Among the location-based experiments, the age group $\geq$10 yrs possibly represents the largest domain shift (on the basis of Pre-Train vs Upper-Bound performances). Similar to the age-based experiment, domain adversarial learning returns the largest relative improvement for kids $\geq$10 yrs. Interestingly, improvements in adversarial learning for kids in 5-10yrs age group are different in Exp 3 and Exp 4. This hints that domain shifts (in this age group) are currently modeled to different extent by GAN and GR, indicating that different modeling techniques should be explored to address this issue. Score fusion performs the best among all the proposed methods, suggesting the presence of complementary information between GAN and GR methods.

As a qualitative analysis, we present TSNE visualizations of the generator outputs for target data from two sessions of Exp 4 in Figure \ref{fig:tsne}. We plot the embeddings before and after GAN training. In both cases, it is evident from the plots that pre-trained embeddings exhibit confusion between child and adult classes, while GAN training increases the discriminative information between them.

\begin{table}[t]
\begin{center}
\caption{Mean F1-score (\%) treating collection center as domain shift}
\label{tab:location_expt}
\resizebox{0.48\textwidth}{!}{
\begin{tabular}{*{7}{c}} \thickhline
\multirow{2}{4em}{\textbf{Systems}}  & \multicolumn{3}{c}{\textbf{Exp 3(\%)}} & \multicolumn{3}{c}{\textbf{Exp 4(\%)}}\\
&   \textbf{$\leq$5yrs}  & \textbf{5-10yrs} & \textbf{$\geq$10yrs}  & \textbf{$\leq$5yrs}  & \textbf{5-10yrs} & \textbf{$\geq1$0yrs} \\ \thickhline
Pre-Train  & 79.55 & 79.23  & 67.69 & 82.12 & 78.16 & 72.68\\
GAN   & 82.14 & 80.32  & 73.32 & 85.03 & 82.32 & 76.72\\
GR & 81.74 & 80.60  & \textbf{73.57} & 84.53 & 82.96 & 76.61\\
Score Fusion & 82.13 & \textbf{80.64} & 73.46 & \textbf{85.21} & \textbf{83.20} & \textbf{76.85}\\
Embed. Fusion & \textbf{82.39} & 80.31 & 73.19 & 82.72 & 82.87 & 75.33\\ \hline
Upper-bound  &  87.72 & 87.56  & 86.74 & 90.67 & 89.47 & 87.80\\\thickhline
\end{tabular}}
\end{center}
\vspace{-3.0ex}
\end{table}

\section{Conclusion}
\label{sec:print}
Previous studies have established the potential of adversarial learning for addressing domain mismatch. In this work, we have applied domain adversarial training to enhance the speaker classification performance in autism diagnosis sessions. We have used 2 different methods (\textit{GAN} and \textit{GR}) for learning domain invariant features, and show that domain adversarial training improves the speaker classification performance by a significant margin. Further, we improved the performance further by fusing at the embedding-level and score-level. 
While our proposed approaches provide improvements over the baseline, the possible upper bound performance implies still significant room for improvement. In the future, we would like to extend adversarial learning to different GAN variants and tasks in the speech pipeline, for example, child ASR.

\bibliographystyle{IEEEbib}
\bibliography{main}

\end{document}